\documentclass[%
aip,
jcp,%
amsmath,amssymb,
reprint%
]{revtex4-1}
\usepackage[latin1]{inputenc}

\usepackage{amsmath}
\usepackage{amsfonts}
\usepackage{amssymb,bm}
\usepackage{graphicx}
\usepackage{physics}
\usepackage{upgreek}
\usepackage[outercaption]{sidecap}   
\usepackage{color}
\usepackage[normalem]{ulem}
\mathchardef\mhyphen="2D
\usepackage{txfonts}

\DeclareMathAlphabet{\pazocal}{OMS}{zplm}{m}{n}

\begin{document}

\def\bra#1{\left<{#1}\right|}
\def\ket#1{\left|{#1}\right>}
\def\expval#1#2{\bra{#2} {#1} \ket{#2}}
\def\mapright#1{\smash{\mathop{\longrightarrow}\limits^{_{_{\phantom{X}}}{#1}_{_{\phantom{X}}}}}}

\title{An improved path-integral method for golden-rule rates}
\author{Joseph E. Lawrence}
\email{joseph.lawrence@chem.ox.ac.uk}
\affiliation{Department of Chemistry, University of Oxford, Physical and Theoretical\\ Chemistry Laboratory, South Parks Road, Oxford, OX1 3QZ, UK}
\author{David E. Manolopoulos}
\affiliation{Department of Chemistry, University of Oxford, Physical and Theoretical\\ Chemistry Laboratory, South Parks Road, Oxford, OX1 3QZ, UK}

\begin{abstract}
We present a simple method for the calculation of reaction rates in the Fermi golden-rule limit, which accurately captures the effects of tunnelling and zero-point energy. The method is based on a modification of the recently proposed golden-rule quantum transition state theory (GR-QTST) of Thapa, Fang and Richardson. While GR-QTST is not size consistent, leading to the possibility of unbounded errors in the rate, our modified method has no such issue and so can be reliably applied to condensed phase systems.  Both methods involve path-integral sampling in a constrained ensemble; the two methods differ, however, in the choice of constraint functional.  We demonstrate numerically that our modified method is as accurate as GR-QTST for the one-dimensional model considered by Thapa and coworkers. We then study a multi-dimensional spin-boson model, for which our method accurately predicts the true quantum rate, while GR-QTST breaks down with an increasing number of boson modes in the discretisation of the spectral density. Our method is able to accurately predict reaction rates in the Marcus inverted regime, without the need for the analytic continuation required by Wolynes theory.
\end{abstract}

\maketitle
\section{Introduction}
There now exist several well established methods which accurately capture the effects of nuclear tunnelling and zero-point energy on electronically adiabatic reaction rates.\cite{Voth89,Craig04,Craig05a,Craig05b,Collepardo08,Boekelheide11,Lawrence20a} However, for reactions which involve an electron transfer the Born-Oppenheimer approximation is usually not applicable and one must include more than one electronic state. The most commonly used approach for understanding and predicting the rates of this type of reaction is Marcus Theory,\cite{Marcus56,Marcus85,Hush61,Hush99} which assumes that the nuclei can be treated classically and that the resulting free energies associated with each electronic state are harmonic. Although it has achieved great success, Marcus theory is limited both by its assumption of harmonic free energies and because it neglects nuclear quantum effects.\cite{Levich70,Kestner74,Ulstrup75,Jortner76,Ulstrup79,Wolynes87} In 1987 Wolynes suggested an approach which aims to solve these problems,\cite{Wolynes87} by taking a saddle point approximation to the exact quantum mechanical flux-flux correlation function in the non-adiabatic (Fermi golden-rule) limit. Wolynes explained that the dominant saddle point occurs on the imaginary-time axis at $t=i\lambda_{\mathrm{sp}}\hbar$, and that his approximation to the rate could therefore be calculated  using imaginary-time path integrals. The resulting theory is straightforward to apply to realistic simulations of non-adiabatic reactions, and it was used in several atomistic studies of electron transfer soon after.\cite{Zheng89,Zheng91,Bader90} 

Despite its success there are several remaining issues with Wolynes theory.  The first is that, although it recovers Marcus theory in the high temperature limit for the spin-boson model, it does not recover the correct (classical golden-rule) expression for the high-temperature rate of an anharmonic reaction.\cite{Richardson15a}  The second issue is that the path-integral representation of the Wolynes rate is only valid when $\lambda_{\mathrm{sp}}$ lies in the interval $[0,\beta]$.\cite{Richardson14,Richardson15a,Lawrence18,Thapa19,Fang19} In the Marcus inverted regime, where the driving force is larger than the reorganisation energy, it turns out that $\lambda_{\mathrm{sp}}<0$. Hence Wolynes theory cannot be directly applied to reactions in this regime. We have shown in a recent paper that it is possible to evaluate the Wolynes rate even when $\lambda_{\mathrm{sp}}<0$, by analytically continuing along the imaginary-time axis.\cite{Lawrence18} However it would clearly be desirable to have a method which avoided the need for this kind of numerical analytic continuation.  The third issue is that for systems with multiple transition states, Wolynes theory can significantly overestimate the rate.\cite{Fang19} This is again because the Wolynes rate must be evaluated at $i\lambda_{\mathrm{sp}}\hbar$. For systems with only one transition state, evaluating the Wolynes expression away from the correct value of $\lambda_{\rm sp}$ leads to a significant overestimation of the rate. Hence when applied to systems with two or more transition states, which when treated separately would have different values of $\lambda_{\mathrm{sp}}$, Wolynes theory is forced to take an intermediate value which overestimates the rate via both channels.\cite{Fang19} 

In the high temperature limit, one can simply use the classical golden-rule expression for the rate.\cite{Chandler98} This is independent of the imaginary time $i\lambda\hbar$ at which it is evaluated, and can therefore be used both in the inverted regime and for systems with multiple transition states.  One might hope that it would be possible to find a quantum generalisation of the classical golden-rule expression that is applicable to lower temperatures and remains independent of the imaginary time at which it is evaluated. Recently Thapa, Fang and Richardson have suggested such a generalisation, which they call golden-rule Quantum Transition State Theory (GR-QTST).\cite{Thapa19,Fang19} This is based on introducing an energy matching constraint which is satisfied by the semiclassical instanton. Unlike Wolynes theory, GR-QTST correctly captures the classical golden-rule rate in the high temperature limit for all systems (with harmonic or anharmonic potentials). For a one dimensional system of linear crossing potentials, it is also exact at all temperatures and is independent of the imaginary time $i\lambda\hbar$ at which it is evaluated. Thapa \emph{et al}.\cite{Thapa19} demonstrated that for other one dimensional problems the GR-QTST rate remains approximately independent of $\lambda$, which enables the calculation of rates in the Marcus inverted regime without the need for analytic continuation (when $\lambda_{\rm sp}<0$ the GR-QTST rate can be calculated using $\lambda=0$.)

Despite these attractive features, GR-QTST has a major drawback which limits its applicability to multidimensional problems: it is not size consistent. The multidimensional generalisation of GR-QTST is affected by adding additional degrees of freedom, even when they are not coupled to the degrees of freedom that participate in the reaction. For a small but finite number of uncoupled degrees of freedom this makes the method more sensitive to the value of $\lambda$ at which the rate is evaluated, which poses problems when $\lambda=0$ is used to calculate rates in the Marcus inverted regime. However, for a realistic condensed phase system, the problem is more severe: the calculation can become dominated by the physically unimportant degrees of freedom, leading to an unbounded error in the rate. This means for example that the results do not converge for a spin-boson model as one increases the number of bath modes in the discretisation of the spectral density, as we shall show below.

In this paper we suggest a modified method which fixes the size consistency problem. Unlike GR-QTST our expression for the rate only involves the diabatic energy gap and projections onto its gradient. This means that adding uncoupled modes does not affect our approximation to the rate. The modified method retains the desirable features of GR-QTST, being exact in the high temperature limit and in one dimension for two linear crossing potentials. For want of a better name we will therefore refer to the method as the \lq\lq Linear Golden-Rule'' (LGR) approximation. We shall show that this method gives rates that are approximately independent of $\lambda$, and hence that it is able to calculate rates deep in the inverted regime (without any analytic continuation). Although LGR is very similar to GR-QTST, we do not base our constraint on an energy matching condition. As such the two methods are not equivalent even in one dimension, except in the high temperature and linear crossing cases. This means that the LGR approximation does not have the same connection to the semiclassical instanton\cite{Miller75,Richardson15a,Richardson15c,Heller20} as GR-QTST, and so it may become less accurate at very low temperatures. This does not however appear to be the case for the problems studied here, which we believe to be representative of typical chemically relevant regimes. 

Section~\ref{Theory_Section} summarises existing theory. Sec.~\ref{GR-QTST_sec} introduces the GR-QTST approximation, and elaborates on its size inconsistency. Sec.~\ref{LGR_sec} introduces the LGR approximation, and explains how it solves the size consistency problem. Sec.~\ref{LGR_Numerical_Implementation} discusses the numerical implementation of the LGR. Sec.~\ref{LGR_Results} presents example applications to two model problems for which exact results are available for comparison: a one dimensional model of electronic pre-dissociation and a multidimensional spin-boson model.  Sec.~\ref{LGR_Conclusion} concludes the paper.

\section{Background Theory} \label{Theory_Section}
\subsection{Exact Theory}
The Hamiltonian for a 2 level system in the diabatic representation can be written in the form
\begin{equation}
\hat{H}=\hat{H}_0\dyad{0}{0}+\hat{H}_1\dyad{1}{1} + \Delta(\dyad{0}{1}+\dyad{1}{0})
\end{equation}
where
\begin{equation}
\hat{H}_i = \sum_{\nu=1}^{f} \frac{\hat{p}^2_\nu}{2m_\nu} + \hat{V}_i(\bm{q})
\end{equation}
is the nuclear Hamiltonian on state $i$, with diabatic potential $\hat{V}_i(\bm{q})$, and $\Delta$ is the electronic coupling. In the following we make the Condon approximation\cite{Nitzan06} and assume that $\Delta$ is independent of the coordinates. Generalisations beyond this are possible but will not be considered here.

The rate constant for transfer from state $\ket{0}$ to state $\ket{1}$ in the non-adiabatic (Fermi golden-rule) limit can be written as the time integral of a flux-flux correlation function\cite{Lax52,Kubo55,Yamamoto60,Miller74,Wolynes87}
\begin{equation}
k = \frac{\Delta^2}{Q_r\hbar^2}\int_{-\infty}^{\infty} c(t+i\lambda\hbar) \mathrm{d}t \label{exact_rate}
\end{equation}
where $Q_r$ is the reactant partition function and
\begin{equation}
c(t) = \tr[e^{-\beta\hat{H}_0}e^{-i\hat{H}_0 t/\hbar}e^{+i \hat{H}_1 t/\hbar}]. \label{Correlation_Function}
\end{equation}
The rate constant in Eq.~(\ref{exact_rate}) is independent of the shift in imaginary time $i\lambda \hbar$; it can be evaluated for any value of $\lambda$ in $[0,\beta]$. This is an important property of the exact rate and as we shall see later it is a desirable property for any approximate theory.

In order to better understand the approximate theories we shall discuss below it is helpful to recast the above expression in terms of the probability distribution
\begin{equation}
\rho_{\lambda}(E) = \frac{1}{2\pi\hbar} \int_{-\infty}^{\infty} \Big\langle e^{-i\hat{H}_0 t/\hbar}e^{+i \hat{H}_1 t/\hbar} \Big\rangle_\lambda e^{-iE t/\hbar}  \mathrm{d}t,
\end{equation}
which is simply a rescaled Fourier transform of $c(t+i\lambda\hbar)$, with
\begin{equation}
\Big\langle e^{-i\hat{H}_0 t/\hbar}e^{+i \hat{H}_1 t/\hbar} \Big\rangle_\lambda = \frac{c(t+i\lambda\hbar)}{c(i\lambda\hbar)}.
\end{equation}
Since the time integral of $c(t+i\lambda\hbar)$ is trivially related to $\rho_{\lambda}(0)$, we can rewrite the rate as
\begin{equation}
k = \frac{2\pi\Delta^2}{Q_r\hbar}\rho_{\lambda}(0)e^{-\beta F(\lambda)},
\end{equation}
where
\begin{equation}
e^{-\beta F(\lambda)}=\tr[e^{-(\beta-\lambda)\hat{H}_0}e^{-\lambda \hat{H}_1}]\label{imaginary_cff}
\end{equation}
is simply the flux-flux correlation function evaluated on the imaginary axis, $c(i\lambda\hbar)$.

Writing the rate in this way also highlights the effect that changing the driving force has on the reaction. To see this we note that introducing an additional bias to products, \ $V_1(\bm{q})\to V_1(\bm{q})-\epsilon$, modifies the correlation function such that
\begin{equation}
\tr[e^{-\beta\hat{H}_0}e^{-i\hat{H}_0 t/\hbar}e^{+i \hat{H}_1 t/\hbar}]\to\tr[e^{-\beta\hat{H}_0}e^{-i\hat{H}_0 t/\hbar}e^{+i \hat{H}_1 t/\hbar}]e^{-i \epsilon t/\hbar}.
\end{equation}
It follows straightforwardly from this that the rate as a function of the driving force, $k(\epsilon)$, is 
\begin{equation}
k(\epsilon) = \frac{2\pi\Delta^2}{Q_r\hbar} \rho_{\lambda}(\epsilon) e^{-\beta F(\lambda)+\lambda\epsilon}. \label{epsilon_dep_rate}
\end{equation}
Hence a knowledge of the distribution calculated with $\epsilon=0$, $\rho_\lambda(E)$ in Eq.~(5),  allows one to calculate the rate at any other driving force.

\subsection{Path-Integral Representation}

Since $e^{-\beta F(\lambda)}$ only involves imaginary-time propagators, it can readily be evaluated using imaginary-time path-integral techniques.\cite{Feynman65,Chandler81,Parrinello84} A standard path-integral discretisation of Eq.~(8) gives\cite{Lawrence18,Cao97}
\begin{equation}
e^{-\beta F(\lambda_l)} = \lim_{n\to \infty} \frac{1}{(2\pi\hbar)^{nf}} \int \mathrm{d}^{nf} \mathbf{p} \int \mathrm{d}^{nf} \mathbf{q}\, e^{-\beta_n H^{(l)}_{n}(\mathbf{p},\mathbf{q}) }.
\end{equation}
where $\beta_n=\beta/n$ and $\lambda_l/\beta=l/n$. Here the ring-polymer Hamiltonian is
\begin{equation}
H^{(l)}_{n}(\mathbf{p},\mathbf{q})\! = h_{n}(\mathbf{p},\mathbf{q})\! +\!  \sum_{j=0}^{l}w_{jl} V_{1}(\bm{q}_j) + \sum_{j=l}^{n} w_{jl} V_{0}(\bm{q}_j)
\end{equation}
where
\begin{equation}
w_{jl} = 
     \begin{cases}
     0 &\quad\text{if }j=l\,\text{  and  }\,l\in\{0,n\} \\
      \frac{1}{2} &\quad\text{if }j\in\{0,l,n\} \text{ and } l\notin\{0,n\}\\
       1 &\quad\text{otherwise} 
     \end{cases} \label{trapezium_weights}
\end{equation}
and 
\begin{equation}
\begin{aligned}
h_{n}(\mathbf{p},\mathbf{q})= \sum_{j=1}^{n}\sum_{\nu=1}^f \bigg[ \frac{p_{j,\nu}^2}{2 m_{\nu}} + \frac{1}{2} m_\nu \omega_n^2\big(q_{j,\nu}-q_{j-1,\nu}\big)^2\bigg],
\end{aligned}
\end{equation}
with $\omega_n=1/\beta_n\hbar$ and $q_{0,\nu}\equiv q_{n,\nu}$.

Note that this is almost the same as the path-integral representation of the partition function for a problem with a single potential energy surface $V(\bm{q})$. The difference is that, instead of the whole imaginary-time path experiencing the same potential, a fraction $\lambda_l/\beta$ of the path experiences the product diabatic potential, $V_1(\bm{q})$, and the remaining fraction $(\beta-\lambda_l)/\beta$ experiences the reactant diabatic potential, $V_0(\bm{q})$. It is straightforward to show from Eq.~(8) that $F'(\lambda)$ can be evaluated in terms of an expectation value in the corresponding ensemble
\begin{equation}
-\beta F'(\lambda_l) = \big\langle V_-\big(\bm{q}_0\big)\big\rangle_{\lambda_l} 
\end{equation}
where $V_{-}(\bm{q})=V_0(\bm{q})-V_1(\bm{q})$ and the expectation value is defined as
\begin{equation}
\langle A(\mathbf{q})\rangle_{\lambda_l} = \lim_{n\to\infty}\frac{\int \mathrm{d}^{nf} \mathbf{p} \int \mathrm{d}^{nf} \mathbf{q}\, e^{-\beta_n H^{(l)}_{n}(\mathbf{p},\mathbf{q}) }A(\mathbf{q})}{\int \mathrm{d}^{nf} \mathbf{p} \int \mathrm{d}^{nf} \mathbf{q}\, e^{-\beta_n H^{(l)}_{n}(\mathbf{p}, \mathbf{q}) }}.
\end{equation}
Hence using imaginary-time path integrals the evaluation of the Boltzmann factor $e^{-\beta F(\lambda)}/Q_r\equiv e^{-\beta[F(\lambda)-F(0)]}$ is straightforward. Unfortunately, however, the evaluation of $\rho_{\lambda}(E)$ in Eq.~(5), which involves real time propagators, is not so simple. 

In order to simplify the discussion of the following methods it is helpful to work with continuous path-integral notation instead of the discrete form. Integrating out the bead momenta in Eq.~(11) one can write 
\begin{equation}
e^{-\beta F(\lambda)} =  \oint\mathcal{D}\bm{q}(\tau) e^{-S_{\lambda}[\bm{q}(\tau)]/\hbar} \label{path_integral}
\end{equation}
where
\begin{equation}
S_{\lambda}[\bm{q}(\tau)] = S^{(\lambda)}_{0}[\bm{q}(\tau)]+S^{(\lambda)}_{1}[\bm{q}(\tau)]
\end{equation}
denotes the Euclidean action with
\begin{subequations}
\begin{align}
S^{(\lambda)}_{0}[\bm{q}(\tau)]&=   \int_{\lambda\hbar}^{\beta\hbar} \frac{1}{2} \dot{\bm{q}}^T(\tau)\bm{M}\dot{\bm{q}}(\tau)+ V_0(\bm{q}(\tau))\,\mathrm{d}\tau\\
S^{(\lambda)}_{1}[\bm{q}(\tau)]&= \int_0^{\lambda\hbar} \frac{1}{2} \dot{\bm{q}}^T(\tau)\bm{M}\dot{\bm{q}}(\tau)+ V_1(\bm{q}(\tau))\,\mathrm{d}\tau. 
\end{align}
\end{subequations}
Here $\bm{q}(\tau)$ is a cyclic path satisfying $\bm{q}(0)\equiv \bm{q}(\beta\hbar)$, $\bm{M}$ is a diagonal mass matrix with diagonal elements $M_{\nu\nu}=m_{\nu}$, and
the factor arising from the integral over bead momenta has become part of the path-integral measure. In this notation, the expectation value of an arbitrary functional of the path is
\begin{equation}
\big\langle A[\bm{q}(\tau)]\big\rangle_\lambda   = \frac{ \oint\mathcal{D}\bm{q}(\tau)\, e^{-S_{\lambda}[\bm{q}(\tau)]/\hbar} A[\bm{q}(\tau)]}{\oint\mathcal{D}\bm{q}(\tau)\, e^{-S_{\lambda}[\bm{q}(\tau)]/\hbar}},
\end{equation}
so Eq.~(15) for example becomes $-\beta F'(\lambda)=\left<V_-(\bm{q}(0))\right>_{\lambda}$.

\subsection{Wolynes Theory}
The Wolynes theory approximation to the non-adiabatic rate constant is normally thought of as a steepest descent approximation to the integral of $c(t+i\lambda\hbar)$ in Eq.~(3). This gives\cite{Wolynes87,Lawrence18,Cao97}
\begin{equation}
k_{\mathrm{WT}}(\lambda_{\mathrm{sp}})=\frac{\Delta^2}{Q_r\hbar}\sqrt{\frac{2\pi}{-\beta F''(\lambda_{\mathrm{sp}})}}e^{-\beta F(\lambda_{\mathrm{sp}})},
\end{equation}
in which $\lambda_{\mathrm{sp}}$ is given by the saddle point condition
\begin{equation}
-\beta F'(\lambda_{\mathrm{sp}}) = 0. \label{Saddle_Point}
\end{equation}

We note here that Wolynes theory can equivalently be thought of as a Gaussian approximation to the distribution $\rho_{\lambda}(E)$ in Eq.~(5) that is constructed so as to reproduce the 0th, 1st and 2nd moments of the exact distribution:
\begin{equation}
\rho_{\mathrm{WT},\lambda}(E) = \sqrt{\frac{1}{2\pi\mu_{2,\lambda}}} \exp(-\frac{(E-\mu_{1,\lambda})^2}{2\mu_{2,\lambda}}),
\end{equation}
where 
\begin{equation}
\mu_{1,\lambda} = \int_{-\infty}^{\infty} E \rho_{\lambda}(E)\, \mathrm{d}E = \beta F'(\lambda)
\end{equation}
and
\begin{equation}
\mu_{2,\lambda} = \int_{-\infty}^{\infty} (E^2-\mu_{1,\lambda}^2) \rho_{\lambda}(E)\, \mathrm{d}E = -\beta F''(\lambda).
\end{equation}
From this perspective, we see that the saddle point condition in Eq.~(22) corresponds to choosing the value of $\lambda$ such that the first moment (or mean) $\mu_{\lambda,1}$ of the distribution $\rho_{\lambda}(E)$ is zero. If the distribution $\rho_{\lambda}(E)$ is singly-peaked, the Gaussian approximation $\rho_{{\rm WT},\lambda}(E)$ will be most accurate near its mean (where $E=\mu_{1,\lambda}$), and less accurate in the tails. Hence requiring that the mean is at zero ensures that $\rho_{{\rm WT},\lambda}(0)$ will be a good approximation to the exact $\rho_{\lambda}(0)$ which appears in the expression for the rate constant in Eq.~(7).

\subsection{High Temperature Limit}
The physical interpretation of the distribution $\rho_{\lambda}(E)$ becomes much clearer in the high temperature limit, where the exact rate tends to the well-known classical golden-rule rate\cite{Chandler98}
\begin{equation}
k_{\mathrm{cl}}=\frac{2\pi\Delta^2}{Q_r\hbar}\big\langle\delta\big(V_-(\bm{q})\big)\big\rangle_{\mathrm{cl},\lambda} e^{-\beta F_{\mathrm{cl}}(\lambda)}, \label{LZ_Rate}
\end{equation}
where
\begin{equation}
e^{-\beta F_{\mathrm{cl}}(\lambda)} = \frac{1}{(2\pi\hbar)^{f}}\!\! \int \mathrm{d}^{f} \bm{p} \!\!\int \mathrm{d}^{f} \bm{q}\, e^{-(\beta-\lambda)H_0(\bm{p},\bm{q})-\lambda H_1(\bm{p},\bm{q})},\! \label{One_Bead_limit}
\end{equation}
with the expectation value taken in the corresponding ensemble 
\begin{equation}
\langle A \rangle_{\mathrm{cl},\lambda} = \frac{\int \mathrm{d}^{f} \bm{p} \int \mathrm{d}^{f} \bm{q} \,  e^{-(\beta-\lambda)H_0(\bm{p},\bm{q})-\lambda H_1(\bm{p},\bm{q})}\,A(\bm{q})}{\int \mathrm{d}^{f} \bm{p} \int \mathrm{d}^{f} \bm{q}\, e^{-(\beta-\lambda)H_0(\bm{p},\bm{q})-\lambda H_1(\bm{p},\bm{q})}} .
\end{equation}
 Hence we see that in this limit the distribution 
 \begin{equation}
 \rho_{\lambda}(E)\to\rho_{\mathrm{cl},\lambda}(E)=\big\langle\delta\big(V_-(\bm{q})+E\big)\big\rangle_{\mathrm{cl},\lambda},
 \end{equation}
such that classically $\rho_{\mathrm{cl},\lambda}(0)$ is just the probability density for the system to be found at the diabatic crossing seam in the $\lambda$ ensemble.
 
 It is clear from these equations that the classical golden-rule rate is entirely {\em independent} of the $\lambda$ at which it is evaluated, (although Eq.~(\ref{One_Bead_limit}) may not be convergent for $\lambda$ outside the interval $[0,\beta]$). This is in stark contrast to the Wolynes rate, which not only fails to give the correct result in the high-temperature limit but must still be evaluated at $\lambda_{\mathrm{sp}}$ in this limit. This also means that classically there is no issue with the evaluation of rates in the inverted regime, where one can simply set $\lambda=0$ in Eq.~(26). It would clearly be desirable to find a path-integral generalisation of this equation that was capable of accurately capturing the effects of tunnelling and zero point energy on the rate. 

\section{GR-QTST} \label{GR-QTST_sec}
\subsection{Formulation}
Recently Thapa, Fang and Richardson have attempted to do precisely this. They have proposed a new method (GR-QTST) for calculating golden-rule rates, which like Wolynes theory involves path-integral sampling.\cite{Thapa19,Fang19} They argued that the rate could be approximated by introducing an energy constraint into Eq.~(\ref{path_integral}), which would be exactly satisfied by the semiclassical instanton.\cite{Richardson15a,Richardson15c,Heller20} Following this logic they suggested the following approximation to the rate\cite{Thapa19}
\begin{equation}
k_{\text{GR-QTST}}(\lambda) = \frac{2\pi\Delta^2}{Q_r\hbar} \bigg\langle\delta\bigg(\frac{2}{3}\mathcal{E}^{(\lambda)}_-[\bm{q}(\tau)]\bigg)\bigg\rangle_\lambda e^{-\beta F(\lambda)},
\end{equation}
where the constraint functional is given by
\begin{equation}
\mathcal{E}^{(\lambda)}_-[\bm{q}(\tau)]=\mathcal{T}_{-}^{(\lambda)}[\bm{q}(\tau)] +  \mathcal{V}_{-}^{(\lambda)}[\bm{q}(\tau)]. \label{GRQTST_constraint}
\end{equation}
Here the first term corresponds to the difference of the virial estimator for the kinetic energy,\cite{Herman82,Tuckerman10} averaged around each segment of the ring-polymer
\begin{equation}
\mathcal{T}_{-}^{(\lambda)}[\bm{q}(\tau)] =\mathcal{T}_{0}^{(\lambda)}[\bm{q}(\tau)]-\mathcal{T}_{1}^{(\lambda)}[\bm{q}(\tau)]
\end{equation}
with
\begin{subequations}
\begin{align}
\mathcal{T}_{0}^{(\lambda)}[\bm{q}(\tau)] = \int_{\lambda\hbar}^{\beta\hbar} \frac{\nabla V_0(\bm{q}(\tau))\dotproduct(\bm{q}(\tau)-\bm{s}(\bm{q}^+))}{2(\beta-\lambda)\hbar}\mathrm{d}\tau\\
\mathcal{T}_{1}^{(\lambda)}[\bm{q}(\tau)]= \int_0^{\lambda\hbar}  \frac{\nabla V_1(\bm{q}(\tau))\dotproduct(\bm{q}(\tau)-\bm{s}(\bm{q}^+))}{2\lambda\hbar}\mathrm{d}\tau
\end{align}
\end{subequations}
and
\begin{equation}
\bm{s}(\bm{q}^+)=\bm{q}^+-\frac{V_-(\bm{q}^+) \nabla V_-(\bm{q}^+)}{|\nabla V_-(\bm{q}^+)|^2}
\end{equation}
with $\bm{q}^+=\frac{1}{2}\big(\bm{q}(0)+\bm{q}(\lambda\hbar)\big)$. The second term in Eq.~(\ref{GRQTST_constraint}) corresponds to the difference between the potential energies average around the two segments 
\begin{equation}
\mathcal{V}_{-}^{(\lambda)}[\bm{q}(\tau)] =\mathcal{V}_{0}^{(\lambda)}[\bm{q}(\tau)]-\mathcal{V}_{1}^{(\lambda)}[\bm{q}(\tau)] \label{V_functionals}
\end{equation}
where
\begin{subequations}
\begin{align}
\mathcal{V}_{0}^{(\lambda)}[\bm{q}(\tau)] = \int_{\lambda\hbar}^{\beta\hbar} \frac{V_0(\bm{q}(\tau))}{(\beta-\lambda)\hbar}\mathrm{d}\tau\\
\mathcal{V}_{1}^{(\lambda)}[\bm{q}(\tau)]= \int_0^{\lambda\hbar}  \frac{V_1(\bm{q}(\tau))}{\lambda\hbar}\mathrm{d}\tau. 
\end{align}
\end{subequations}

As with Wolynes theory, GR-QTST is expected to be most accurate when $\rho_{\lambda}(0)$ is near the peak of the distribution and so ideally $k_{\text{GR-QTST}}(\lambda)$ should be evaluated at $\lambda=\lambda_{\mathrm{sp}}$. However, when $\lambda_{\mathrm{sp}}$ is outside the interval $[0,\beta]$, such that evaluation at $\lambda_{\mathrm{sp}}$ is not possible, Thapa \emph{et al.} suggest evaluating the rate at the end point closest to $\lambda_{\mathrm{sp}}$.\cite{Thapa19}

In low dimensions Thapa \emph{et al.} found their method gave very good agreement with exact quantum mechanical golden-rule rates, and that since $k_{\text{GR-QTST}}(\lambda)$ is approximately independent of $\lambda$ it gives accurate predictions of rates in the Marcus inverted regime and for systems with multiple transition states.\cite{Thapa19,Fang19} Unfortunately, however, their method is not size consistent. Although this size inconsistency has been discussed previously by Richardson and coworkers,\cite{Thapa19,Fang19} its origins and implications have not yet been fully explored.

\subsection{Size Inconsistency}
To explain the lack of size consistency in GR-QTST, we shall consider a system for which the coordinates can be separated into two uncoupled sets $\bm{q}_a$ and $\bm{q}_b$, and for which only $\bm{q}_a$ are coupled to the diabatic states. We note that this is not a physically unreasonable model as in a real atomistic simulation there may be many degrees of freedom which are essentially uncoupled from the non-adiabatic reaction of interest. The diabatic potentials for such a system can be written as 
\begin{subequations}
\begin{align}
V_0(\bm{q}) &= U_{0,a}(\bm{q}_a) + U_b(\bm{q}_{b})  \\
V_1(\bm{q}) &= U_{1,a}(\bm{q}_a) + U_b(\bm{q}_{b}),
\end{align}
\end{subequations}
with corresponding diabatic Hamiltonians
\begin{subequations}
\begin{align}
\hat{H}_0 &= \hat{H}_{0,a} + \hat{H}_b  \\
\hat{H}_1 &= \hat{H}_{1,a} + \hat{H}_b.
\end{align}
\end{subequations}
It is clear physically that the rate of transfer from state $\ket{0}$ to state $\ket{1}$ is completely independent of the ``$b$'' degrees of freedom. This can be confirmed by considering Eq.~(\ref{exact_rate}) and noting that since $[\hat{H}_{i,a},\hat{H}_b]=0$
\begin{equation}
\begin{aligned}
c(t) &=  \tr[e^{-\beta\hat{H}_0}e^{-i\hat{H}_0 t/\hbar}e^{+i \hat{H}_1 t/\hbar}] \\
&= \tr_{a}\Big[e^{-\beta\hat{H}_{0,a}}e^{-i\hat{H}_{0,a} t/\hbar}e^{+i \hat{H}_{1,a} t/\hbar}\Big]\tr_b\Big[e^{-\beta\hat{H}_{b}}\Big]
\end{aligned}
\end{equation}
such that the ``$b$'' dependent terms cancel exactly with those in the reactant partition function
\begin{equation}
Q_r = \tr_{a}\Big[e^{-\beta\hat{H}_{0,a}}\Big]\tr_b\Big[e^{-\beta\hat{H}_{b}}\Big].
\end{equation}

The GR-QTST expression for the rate, however, does not have this property. To see this we first define the probability distribution
\begin{equation}
p_{\lambda}(E) = \bigg\langle\delta\bigg(\frac{2}{3}\mathcal{E}^{(\lambda)}_-[\bm{q}(\tau)]-E\bigg)\bigg\rangle_\lambda
\end{equation}
which is equivalent to the pre-exponential term in $k_{\text{GR-QTST}}(\lambda)$ when evaluated at $E=0$. (Note that, due to the $\epsilon$ dependence of $\bm{s}(\bm{q}^{+})$ in Eq.~(34), this is not the same as $\rho_{\text{GR-QTST},\lambda}(E)$). Then noting that that the constraint functional can be separated into two parts, each of which only depends on one of the two uncoupled sets of degrees of freedom, we can write
\begin{equation}
\mathcal{E}^{(\lambda)}_-[\bm{q}(\tau)] = \mathcal{E}^{(\lambda)}_{-,a}[\bm{q}_a(\tau)]+\mathcal{E}^{(\lambda)}_{-,b}[\bm{q}_b(\tau)].
\end{equation}
It then follows straightforwardly that 
\begin{equation}
p_{\lambda}(E)  = \int_{-\infty}^{\infty} p_{a,\lambda}(E-E')p_{b,\lambda}(E')\,{\rm d}E',
\end{equation}
where
\begin{equation}
p_{\alpha,\lambda}(E) = \bigg<\delta\bigg({2\over 3}\mathcal{E}^{(\lambda)}_{-,\alpha}[\bm{q}_{\alpha}(\tau)]-E\bigg)\bigg>_\lambda
\end{equation}
with $\alpha=a\text{ or } b$. 

If it were true that $\mathcal{E}_{-,b}^{(\lambda)}[\bm{q}_{b}(\tau)]=0$, then $p_{b,\lambda}(E)$ would simply be $\delta(E)$, and GR-QTST would correctly predict that the spectator degrees of freedom did not affect the rate.  However, this is not in general the case as the instantaneous quantum fluctuations in the virial energy estimators for these degrees of freedom averaged around the two segments of the ring polymer do not perfectly cancel. It follows that $p_{b,\lambda}(E)$ is a distribution with non-zero standard deviation. Hence, when it is convoluted with $p_{a,\lambda}(E)$, the resulting distribution will not in general satisfy $p_{\lambda}(0)= p_{a,\lambda}(0)$.  This problem becomes worse with increasing system size, as the variance of the distribution $p_{b,\lambda}(E)$ grows linearly with the number of uncoupled degrees of freedom.  While the fluctuations may be small when there are only a few uncoupled degrees of freedom, they will clearly eventually come to dominate the rate.
Although Richardson and coworkers have discussed the size inconsistency of GR-QTST previously,\cite{Thapa19,Fang19} the analysis presented here clarifies under what circumstances the GR-QTST rate will break down.  In particular we can see that the rate will be significantly affected when the variance due to fluctuations in degrees of freedom uncoupled to the reaction becomes comparable to the variance of the exact distribution. As the variance of the exact distribution is on the order of $\Lambda/\beta$, where $\Lambda$ is the Marcus reorganisation energy, it is simple to assess whether uncoupled degrees of freedom are likely to dominate the rate. The size inconsistency will be most pronounced in the tails of the distribution, and for singly peaked $\rho_{\lambda}(E)$ the error is thus expected to be smallest when the GR-QTST rate is evaluated at $\lambda_{\mathrm{sp}}$.

 We shall show in a forthcoming paper that, for a realistic atomistic simulation of aqueous ferrous-ferric electron transfer, the solvent degrees of freedom that are uncoupled from the reaction do indeed dominate the GR-QTST approximation to $\rho_\lambda(E)$, leading to a spurious prediction of the rate.\cite{Lawrence20c} Unfortunately, even in relatively small systems, where all degrees of freedom are significantly coupled to the diabatic electronic states, the lack of size consistency in GR-QTST can lead to large errors in the predicted rates, as we shall illustrate for a spin boson model in Sec.~\ref{LGR_Results}.  

\section{An Improved Method} \label{LGR_sec}
\subsection{Linear crossing potentials}
GR-QTST does have some desirable features which would make it quite attractive if it were not for its size inconsistency. We believe the most important of these is that, for a one-dimensional system of linear crossing potentials,
\begin{subequations}
\begin{align}
V_0(q) &= \kappa_0 (q-q^\ddagger) + V^\ddagger \\
V_1(q) &= \kappa_1 (q-q^\ddagger) + V^\ddagger, 
\end{align}
\end{subequations} 
the method gives the exact quantum mechanical rate constant\cite{Richardson15a}
\begin{equation}
k Q_r=\sqrt{\frac{2\pi m}{\beta \hbar^2}} \frac{\Delta^2}{\hbar|\kappa_0-\kappa_1|}e^{-\beta V^\ddagger}\exp(\frac{\beta^3\hbar^2\kappa_0^2\kappa_1^2}{24m(\kappa_0-\kappa_1)^2}),
\end{equation}
independent of the choice of $\lambda$. In order to show this, Thapa \emph{et al.}\cite{Thapa19} demonstrated that for this system the exact rate can be written in terms of an imaginary-time path integral as
\begin{equation}
k = \frac{2\pi\Delta^2}{\hbar Q_r} \oint\mathcal{D}q(\tau)\, e^{-S_\lambda[q(\tau)]/\hbar} \delta(\mathcal{V}^{(\lambda)}_{-}[q(\tau)]) \label{linear_rate}
\end{equation}
where
\begin{equation}
\mathcal{V}_{-}^{(\lambda)}[q(\tau)] = \mathcal{V}_{0}^{(\lambda)}[q(\tau)]-\mathcal{V}_{1}^{(\lambda)}[q(\tau)]
\end{equation}
with
\begin{subequations}
\begin{align}
\mathcal{V}_{0}^{(\lambda)}[q(\tau)] &= V_0(\tilde{q}) + \int_{\lambda\hbar}^{\beta\hbar} \frac{\kappa_0(q(\tau)-\tilde{q})}{(\beta-\lambda)\hbar}\mathrm{d}\tau\\
\mathcal{V}_{1}^{(\lambda)}[q(\tau)]&= V_1(\tilde{q}) + \int_0^{\lambda\hbar}  \frac{\kappa_1(q(\tau)-\tilde{q})}{\lambda\hbar}\mathrm{d}\tau, 
\end{align} \label{linear_functionals}
\end{subequations}
for any choice of $\tilde{q}$. Instead of merely being a step on the way to showing that GR-QTST is exact in the linear case, we believe that this result is instead fundamental to the success of the method for more general one-dimensional problems.
 
\subsection{Linear Golden-Rule Approximation}
Following this perspective, let us now suggest a modified method which directly generalises Eqs.~(\ref{linear_rate}) to (\ref{linear_functionals}) for multidimensional and non-linear potentials. The most obvious generalisation of Eqs.~(48) and~(49) would be to use the difference between diabatic potential energies averaged around each segment of the imaginary-time path,
\begin{equation}
\mathcal{V}_{-}^{(\lambda)}[\bm{q}(\tau)] =  \int_{\lambda\hbar}^{\beta\hbar} \frac{V_0(\bm{q}(\tau))}{(\beta-\lambda)\hbar}\mathrm{d}\tau-\int_{0}^{\lambda\hbar} \frac{V_1(\bm{q}(\tau))}{\lambda\hbar}\mathrm{d}\tau. \label{primitive_diabatic_gap}
\end{equation}
However, for the reasons we have already discussed, this is not size consistent. For a two dimensional system described by the diabats
\begin{subequations}
\begin{align}
&V_i(\bm{q}) = U_{i,a}(q_a) + U_b(q_{b})  \\
&U_{i,a}(q_a) = \kappa_i (q-q^\ddagger) + V^\ddagger,
\end{align}
\end{subequations}
it is clear that the multidimensional generalisation of Eqs.~(48) and (49) that avoids the size consistency problem would need to be equivalent to 
\begin{equation}
\mathcal{U}_{-}^{(\lambda)}[q_a(\tau)] =  \int_{\lambda\hbar}^{\beta\hbar} \frac{U_{0,a}(q_a(\tau))}{(\beta-\lambda)\hbar}\mathrm{d}\tau-\int_{0}^{\lambda\hbar} \frac{U_{1,a}(q_a(\tau))}{\lambda\hbar}\mathrm{d}\tau,
\end{equation}
rather than Eq.~(\ref{primitive_diabatic_gap}). This suggests that we would ideally like to ``project out'' degrees of freedom uncoupled to $V_{-}(\bm{q})$. In this simple example this can clearly be achieved by letting
\begin{equation}
\mathcal{U}_{-}^{(\lambda)}[\bm{q}(\tau)] = \int_{\lambda\hbar}^{\beta\hbar} \frac{V_0(\tau,\tau')}{(\beta-\lambda)\hbar}\mathrm{d}\tau-\int_{0}^{\lambda\hbar} \frac{V_{1}(\tau,\tau')}{\lambda\hbar}\mathrm{d}\tau
\end{equation}
where
\begin{equation}
V_i(\tau,\tau') =  V_i(q_a(\tau),q_b(\tau')),
\end{equation}
i.e., by fixing degrees of freedom orthogonal to the diabatic energy gap coordinate at some specified imaginary time $i\tau'$. 

 In order to generalise this to more complex multidimensional systems we first define the local diabatic energy gap coordinate as
\begin{equation}
x_{\tau'}(\tau) = V_{-}(\bm{q}(\tau')) + \nabla V_{-}(\bm{q}(\tau'))\dotproduct(\bm{q}(\tau)-\bm{q}(\tau')),
\end{equation}
and then consider changes in the diabatic potentials as a function of $x_{\tau'}(\tau)$ whilst keeping orthogonal degrees of freedom fixed at $\tau'$,
\begin{equation}
V_i(\tau,\tau') = V_i(\bm{q}(\tau')) + \frac{\partial V_i}{\partial x_{\tau'}}\bigg|_{\tau'} \big(x_{\tau'}(\tau)-x_{\tau'}(\tau')\big) + \dots .
\end{equation}  
In order to obtain a practical expression for the constraint functional we assume that along the imaginary-time path the diabatic potential can be treated as a harmonic function of $x_{\tau'}$, such that
\begin{equation}
\frac{\partial^2 V_i}{\partial x_{\tau'}^2}\bigg|_{\tau'} \simeq \frac{\displaystyle{\frac{\partial V_i}{\partial x_{\tau'}}\Big|_{\tau}-\frac{\partial V_i}{\partial x_{\tau'}}\Big|_{\tau'}}}{x_{\tau'}(\tau)-x_{\tau'}(\tau')},
\end{equation}
and 
\begin{equation}
V_i(\tau,\tau') \simeq V_i(\bm{q}(\tau')) +\frac{\displaystyle{\frac{\partial V_i}{\partial x_{\tau'}}\Big|_{\tau}+\frac{\partial V_i}{\partial x_{\tau'}}\Big|_{\tau'}}}{2} \big(x_{\tau'}(\tau)-x_{\tau'}(\tau')\big). \label{harmonic_expansion}
\end{equation}
This can then be rewritten explicitly in terms of the original coordinates by noting that 
\begin{equation}
x_{\tau'}(\tau)-x_{\tau'}(\tau') =  \nabla V_{-}(\bm{q}(\tau'))\dotproduct(\bm{q}(\tau)-\bm{q}(\tau'))
\end{equation}
and
\begin{equation}
\frac{\partial V_i}{\partial x_{\tau'}}\bigg|_{\tau} = \frac{\nabla V_i(\bm{q}(\tau))\dotproduct\nabla V_{-}(\bm{q}(\tau'))}{|\nabla V_{-}(\bm{q}(\tau'))|^2},
\end{equation}
which having defined the projected diabatic gradients
\begin{equation}
\bm{\kappa}_{i,\tau'}(\tau) = \frac{\nabla V_i(\bm{q}(\tau))\dotproduct\nabla V_{-}(\bm{q}(\tau'))}{|\nabla V_{-}(\bm{q}(\tau'))|^2} \nabla V_{-}(\bm{q}(\tau'))
\end{equation}
allows us to rewrite Eq.~(\ref{harmonic_expansion}) in the form
\begin{equation}
V_i(\tau,\tau') \simeq V_i(\bm{q}(\tau')) +\frac{\bm{\kappa}_{i,\tau'}(\tau)+\bm{\kappa}_{i,\tau'}(\tau')}{2}\dotproduct \big(\bm{q}(\tau)-\bm{q}(\tau')\big).
\end{equation}

It is clear that in general this approach depends on the value of $\tau'$ about which the expansion is taken.  We suggest averaging $\tau'$ over the two hopping times, $\lambda\hbar$ and $\beta\hbar$, which we have found generally gives the most accurate results.  Our approximation to the rate can then be written as
\begin{equation}
k_{\mathrm{LGR}}(\lambda) = \frac{2\pi\Delta^2}{\hbar Q_r} \big\langle\delta\big(\bar{V}^{(\lambda)}_{-}[\bm{q}(\tau)]+\bar{\mathcal{K}}^{(\lambda)}_{-}[\bm{q}(\tau)]\big)\big\rangle_\lambda e^{-\beta F(\lambda)}, \label{our_method}
\end{equation}
in which the argument of the delta function consists of two terms. The first of these is simply the diabatic energy gap averaged  over the two bridging beads,
\begin{equation}
\bar{V}^{(\lambda)}_{-}[\bm{q}(\tau)]= \frac{V_{-}(\bm{q}(\lambda\hbar))+V_{-}(\bm{q}(\beta\hbar))}{2},  \label{diabatic_difference_functional}
\end{equation}
and the second is a gradient-based correction of the form
\begin{equation}
\bar{\mathcal{K}}^{(\lambda)}_{-}[\bm{q}(\tau)] = \bar{\mathcal{K}}^{(\lambda)}_{0}[\bm{q}(\tau)]-\bar{\mathcal{K}}^{(\lambda)}_{1}[\bm{q}(\tau)]
\end{equation} 
where
\begin{equation}
\bar{\mathcal{K}}^{(\lambda)}_{i}[\bm{q}(\tau)] = \frac{\mathcal{K}^{(\lambda)}_{i,\lambda\hbar}[\bm{q}(\tau)]+\mathcal{K}^{(\lambda)}_{i,\beta\hbar}[\bm{q}(\tau)]}{2}
\end{equation}
with
\begin{subequations} \label{Correction_terms}
\begin{align}
 \mathcal{K}^{(\lambda)}_{0,\tau'}[\bm{q}(\tau)]&=\! \int_{\lambda\hbar}^{\beta\hbar}\frac{\bar{\bm{\kappa}}_0(\tau,\tau')\dotproduct\big(\bm{q}(\tau)-\bm{q}(\tau')\big)}{(\beta-\lambda)\hbar}  \mathrm{d}\tau \\
 \mathcal{K}^{(\lambda)}_{1,\tau'}[\bm{q}(\tau)]&=\! \int_{0}^{\lambda\hbar}\frac{\bar{\bm{\kappa}}_1(\tau,\tau')\dotproduct\big(\bm{q}(\tau)-\bm{q}(\tau')\big)}{\lambda\hbar}  \mathrm{d}\tau
\end{align} 
\end{subequations}
and 
\begin{equation}
\bar{\bm{\kappa}}_i(\tau,\tau') = \frac{\bm{\kappa}_{i,\tau'}(\tau) + \bm{\kappa}_{i,\tau'}(\tau')}{2}.\label{kappa_bar_def}
\end{equation}

Since this method is based directly on the exact result in the linear case and only involves the gradients of the diabatic potentials we call it the linear golden-rule (LGR) approximation. In view of the assumption we have made to obtain Eqs.~(57) and~(58) we expect it will be most accurate when the gradients of the diabatic potentials do not change dramatically around the imaginary-time path. The LGR approximation does not have the same connection to the semiclassical instanton that Thapa \emph{et al.}\cite{Thapa19} chose to prioritise in their definition of GR-QTST, and hence is likely to be less accurate for low dimensional systems at low temperatures. It is however clear by construction that it retains the property of being exact for a one dimensional system of two linear crossing potentials, as well as reducing to Eq.~(\ref{LZ_Rate}) in the high temperature limit (where $|\bm{q}(\tau)-\bm{q}(\tau')|\to0$ and $\bar{\mathcal{K}}^{(\lambda)}_{-}[\bm{q}(\tau)]\to 0$). In addition to this we note that, unlike GR-QTST, the LGR approximation to the distribution $\rho_{\lambda}(E)$,
\begin{equation}
\rho_{\mathrm{LGR},\lambda}(E) =  \big\langle\delta\big(\bar{V}^{(\lambda)}_{-}[\bm{q}(\tau)]+\bar{\mathcal{K}}^{(\lambda)}_{-}[\bm{q}(\tau)]+E\big)\big\rangle_\lambda, \label{LGR_rho}
\end{equation}
correctly integrates to one, and hence is a true distribution.  

Except in the linear case and the high temperature limit, $k_{\mathrm{LGR}}(\lambda)$ will not be completely independent of $\lambda$. Hence, just as with Wolynes theory and GR-QTST, we expect that $k_{\mathrm{LGR}}(\lambda)$ will be most accurate when evaluated at the saddle point $\lambda_{\mathrm{sp}}$. As with GR-QTST, we thus suggest that the LGR rate is evaluated at $\lambda_{\mathrm{sp}}$ or at the closest end point when $\lambda_{\mathrm{sp}}$ falls outside the range $[0,\beta]$.

We note that the first of the two terms appearing in the delta function in Eq.~(\ref{LGR_rho}) is precisely the same as that which arrises from making the static approximation,\cite{Sparpaglione88}
\begin{equation}
e^{-i\hat{H}_0t/2\hbar}e^{+i\hat{H}_1t/2\hbar} \simeq e^{-i\hat{V}_-t/2\hbar} 
\end{equation}
symmetrically in Eq.~(\ref{exact_rate}) to give
\begin{equation}
c(t+i\lambda\hbar)\simeq\tr[e^{-(\beta-\lambda)\hat{H}_0}e^{-i\hat{V}_-t/2\hbar}e^{-\lambda\hat{H}_1}e^{-i\hat{V}_-t/2\hbar}].
\end{equation}
Although it is exact at high temperature, the static approximation is generally only valid at short time. Hence at low temperatures the static approximation to the distribution,
\begin{equation}
\rho_{\mathrm{st},\lambda}(E) = \big\langle\delta\big(\bar{V}^{(\lambda)}_{-}[\bm{q}(\tau)]+E\big)\big\rangle_\lambda,
\end{equation}
is typically not accurate in the tails of the distribution. The second term in the argument of the delta function in Eq.~(\ref{LGR_rho}), $\bar{\mathcal{K}}^{(\lambda)}_{-}[\bm{q}(\tau)]$, acts to correct the static approximation to the tails of the distribution, and makes the rate approximately independent of $\lambda$. The LGR approximation therefore combines the exact quantum statistics of the system with a local linear approximation to predict how tunnelling and zero-point energy will affect the rate.



\subsection{Size Consistency}
The main theoretical improvement of the LGR approximation over GR-QTST is that it is not affected by adding uncoupled modes to the system. To highlight this we shall again consider a system with diabatic potentials
\begin{subequations}
\begin{align}
V_0(\bm{q}) &= U_{0,a}(\bm{q}_a) + U_b(\bm{q}_{b})  \\
V_1(\bm{q}) &= U_{1,a}(\bm{q}_a) + U_b(\bm{q}_{b}),
\end{align}
\end{subequations}
in which the ``\emph{b}'' degrees of freedom are not coupled (directly or indirectly) to the diabatic state and hence should have no effect on the rate. Since the diabatic energy gap is then (by definition) independent of $\bm{q}_b$, 
\begin{equation}
V_-(\bm{q}) = U_{0,a}(\bm{q}_a) - U_{1,a}(\bm{q}_a),
\end{equation} 
it follows straightforwardly that so too is its average around the imaginary-time path, $\bar{V}^{(\lambda)}_{-}[\bm{q}(\tau)]=\bar{V}^{(\lambda)}_{-}[\bm{q}_a(\tau)]$. For the correction term we note that since the derivative of the energy gap is independent of $\bm{q}_b$ it follows simply that both
\begin{equation}
\nabla V_i(\bm{q}(\tau)) \dotproduct \nabla V_-(\bm{q}(\tau')) = \nabla U_{i,a}(\bm{q}_a(\tau)) \dotproduct \nabla V_-(\bm{q}_a(\tau')),
\end{equation} 
and
\begin{equation}
\nabla V_{-}(\bm{q}(\tau'))\dotproduct (\bm{q}(\tau)-\bm{q}(\tau')) = \nabla V_{-}(\bm{q}_a(\tau'))\dotproduct (\bm{q}_a(\tau)-\bm{q}_a(\tau')),
\end{equation}
 and hence that the correction term, $\bar{\mathcal{K}}^{(\lambda)}_{-}[\bm{q}(\tau)]=\bar{\mathcal{K}}^{(\lambda)}_{-}[\bm{q}_a(\tau)]$, is also independent of the uncoupled modes. (Note that the gradient operator in Eqs.~(75) and (76) is $\nabla=(\partial/\partial \bm{q}_a,\partial/\partial \bm{q}_b)$, and that cross terms between the \lq\lq $a$" and \lq\lq $b$" coordinates do not contribute to the dot products.) Hence, our method does not suffer the same size consistency issue as GR-QTST. This key difference arrises because of the projection of $\nabla V_i$ onto $\nabla V_-$, which ensures that degrees of freedom which are not coupled to $V_-(\bm{q})$ do not enter the constraint functional. 

\section{Numerical Implementation} \label{LGR_Numerical_Implementation}
There are clearly two components to evaluating $k_{\mathrm{LGR}}(\lambda)$, the first of which is the evaluation of the Boltzmann factor, $e^{-\beta F(\lambda)}/Q_r$. This is common to all of the methods we have discussed: Wolynes Theory, GR-QTST and LGR.  Since
\begin{equation}
{e^{-\beta F(\lambda)}\over Q_r} = e^{-\beta[F(\lambda)-F(0)]} = e^{-\beta\int_0^{\lambda} F'(\lambda')\,{\rm d}\lambda'},
\end{equation}
and since $-\beta F'(\lambda_l)$ is given by the simple path-integral expectation value in Eq.~(15), 
this factor is straightforward to calculate by evaluating $-\beta F'(\lambda_l)$ on a grid of points and doing a numerical integration. We would also reiterate that one then immediately has access to $e^{-\beta F(\lambda)}$ for any modified system with a different driving force, $V_1\to V_1-\epsilon$, since $e^{-\beta F(\lambda)}\to e^{-\beta F(\lambda)+\lambda \epsilon}$. Hence one calculation gives $F(\lambda)$ for all driving forces. 
 
 The second component in $k_{\mathrm{LGR}}(\lambda)$ is the probability density, $\rho_{\mathrm{LGR},\lambda}(0)$. Since $\rho_{\lambda_{\mathrm{sp}}}(E)$ tends to be peaked around $E=0$, this can be evaluated at the same time as $-\beta F'(\lambda_l)$ by histogramming the the $n$ bead discretisation of $\bar{V}^{(\lambda)}_{-}[\bm{q}(\tau)]+\bar{\mathcal{K}}^{(\lambda)}_{-}[\bm{q}(\tau)]$ to give $\rho_{\mathrm{LGR},\lambda_l}(E)$. Hence in the normal regime one needs to do no additional calculations beyond those required by Wolynes theory.\cite{Wolynes87,Lawrence18} However, in the inverted regime, for reactions which are sufficiently activated, direct evaluation of $\rho_{\mathrm{LGR},\lambda_l=0}(0)$ in this way  will be insufficient. Instead one will need to use an enhanced sampling technique to sample configurations around the crossing seam $V_0(\bm{q})=V_1(\bm{q})$. This added complexity is mitigated by the fact that there is no need in this case to do any simulations for values of $\lambda_l>0$, and the fact that one immediately knows the Boltzmann factor $e^{-\beta F(0)}/Q_r=1$. Hence, for a similar activation energy, the inverted regime does not introduce any greater simulation effort than that required in the normal regime. Finally, we reiterate that, just as with the Boltzmann factor, having calculated the distribution $\rho_{\mathrm{LGR},\lambda}(E)$ for one value of the driving force, it is readily available for any other. This is because $V_1\to V_1-\epsilon$ simply changes the distribution to $\rho_{\mathrm{LGR},\lambda}(E)\to\rho_{\mathrm{LGR},\lambda}(E+\epsilon)$.
 
\section{Results and Discussion} \label{LGR_Results}
In order to test the LGR approximation we have considered two different model problems for which exact results are available for comparison. The first of these is the one dimensional electronic pre-dissociation model considered by Thapa \emph{et al.} in their first paper introducing GR-QTST.\cite{Thapa19}   The second is a multidimensional discretisation of a spin boson model with an exponentially damped Ohmic spectral density, with parameters chosen to provide a demanding test case exhibiting significant nuclear quantum effects.  
  
\subsection{One-dimensional pre-dissociation model}
The first model we consider is the pre-dissociation model originally introduced by Richardson and Thoss to demonstrate the oscillatory nature of flux-flux correlation functions.\cite{Richardson14} This model was also considered by Thapa \emph{et al.}\cite{Thapa19} in their paper introducing GR-QTST, and for ease of comparison with their results we consider the same parameter regimes as they did in their paper. 

For this system the diabatic potentials take the form
\begin{equation}
V_0(q) = \frac{1}{2} \omega^2 q^2
\end{equation}
and
\begin{equation}
V_1(q) = D e^{-2\alpha(q-\xi)} - \epsilon.
\end{equation} 
The model parameters are $m=1$, $\omega=1$, $D=2$, $\alpha=0.2$, and $\xi=5$, in units where $\hbar=1$, and the calculations are performed over a range of values of $\beta$ and $\epsilon$. We define the reorganisation energy for this problem as $\Lambda=De^{2\alpha\xi}$. 

\begin{figure}[t]
 \resizebox{1.0\columnwidth}{!} {\includegraphics{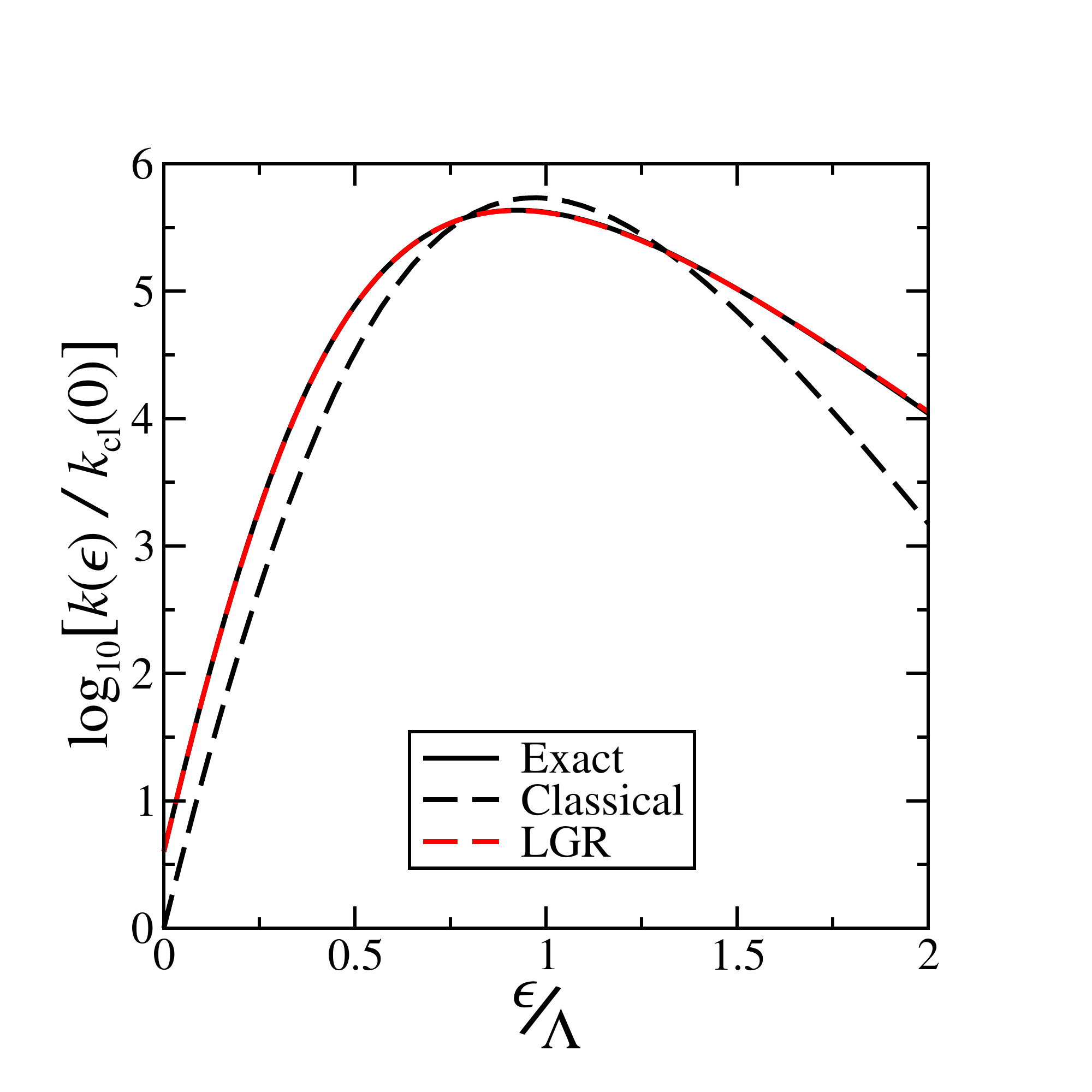}}
 \centering
 \caption{Rate constants for the electronic pre-dissociation model at $\beta=3$, relative to the classical result at $\epsilon=0$ . The exact results were calculated using Eq.~(\ref{DVR}) and the LGR results as described in Sec.~\ref{LGR_Numerical_Implementation}, using 256 ring polymer beads.}
 \label{Pre-Dis_Epsilon}
 \end{figure}
 
    \begin{figure}[t]
 \resizebox{1.0\columnwidth}{!} {\includegraphics{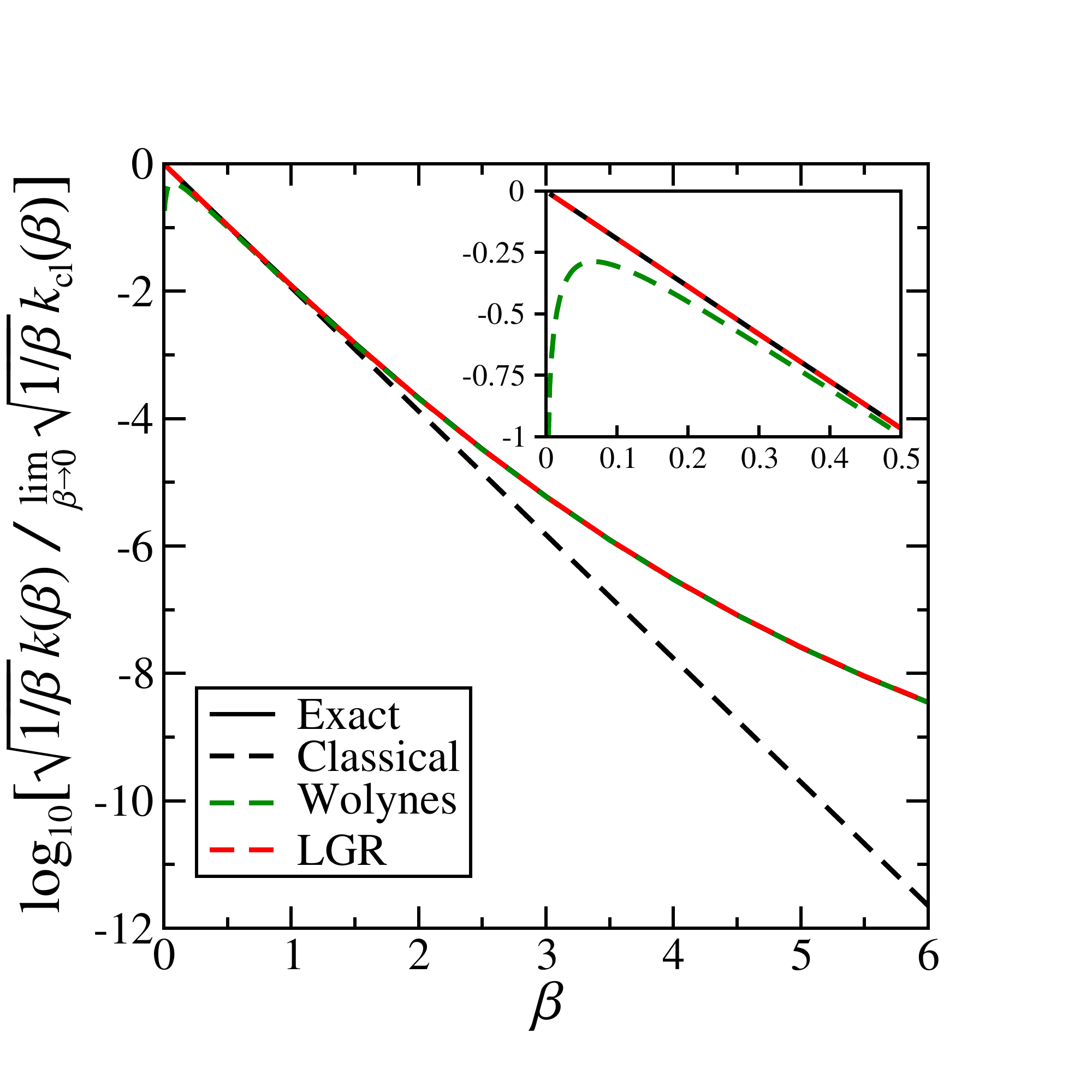}}
 \centering
 \caption{Rate constants as a function of inverse temperature for the pre-dissociation model, with $\epsilon=0$. The exact results were calculated using Eq.~(\ref{DVR}). The results for Wolynes Theory and LGR were calculated using 512 beads at the lowest temperature. The inset shows the break down of Wolynes theory at high temperature. The GR-QTST results are not included as they are essentially indistinguishable from the LGR rates on this plot. They can be found in Ref.~\citenum{Thapa19}. }
 \label{TempDep1}
 \end{figure}

In order to highlight the relative importance of nuclear quantum effects in different regimes, we shall compare our results with the classical rate in Eq.~(\ref{LZ_Rate}). For this simple one dimensional model this can be expressed as\cite{Nitzan06}
\begin{equation}
k_{\text{cl}} = \frac{\Delta^2}{\hbar}\sqrt{2\pi\beta m \omega^2}\frac{e^{-\beta V_0(q^\ddagger)}}{|V'_-(q^{\ddagger})|},
\end{equation}
where $q^{\ddagger}$ satisfies the equation $V_0(q^\ddagger)=V_1(q^\ddagger)$. The exact quantum mechanical rate
\begin{equation}
k=-\frac{2\Delta^2}{\hbar Q_r} \sum_j e^{-\beta E_j} \Im \bra{j}\hat{G}^+_1(E_j)\ket{j}, \label{DVR}
\end{equation} 
with
\begin{equation}
Q_r = \sum_j e^{-\beta E_j}
\end{equation}
where
\begin{equation}
\hat{H}_0\ket{j} = E_j\ket{j}
\end{equation}
and
\begin{equation}
\hat{G}_1^+(E_j) = \lim_{\eta\to0^+} \big(E_j+i\eta-\hat{H}_1\big)^{-1},
\end{equation}
was calculated using a Lobatto shape function discrete variable representation.\cite{Manolopoulos88}

Figure \ref{Pre-Dis_Epsilon} compares the rate as a function of the driving force for this model at a fixed temperature, with all rates plotted relative to the classical rate at zero driving force. We do not include the GR-QTST results in Fig.~\ref{Pre-Dis_Epsilon} as they are the same as the LGR rates to graphical accuracy. The LGR rates are also almost graphically indistinguishable from the exact rates for this problem, both in the normal regime and all the way out to $\epsilon=2\Lambda$ in the inverted regime. The accuracy of the method in the inverted regime is particularly pleasing given the simplicity with which the LGR calculation is performed, in particular when compared with the approach of Ref.~\citenum{Lawrence18}, which requires numerical analytic continuation to obtain the Wolynes theory rate in the inverted regime. Comparison with the classical rate shows that at this temperature nuclear quantum effects only have a moderate effect on the rate, with the largest difference, occurring at $\epsilon=2\Lambda$, corresponding to less than an order of magnitude speed up relative to the classical rate.

Figure \ref{TempDep1} shows how the rate constants vary as a function of temperature at zero driving force, $\epsilon=0$. We see clearly the break down of Wolynes theory at high temperature, as is highlighted in the inset of the figure. GR-QTST and LGR do not suffer from this problem, and instead correctly recover the exact rate in the high temperature limit. At lower temperatures we see that both LGR and Wolynes theory are graphically indistinguishable from the exact rate (which is hidden under the LGR and Wolynes curves in the figure). We find that the LGR rates are very accurate for this problem, with an error of less than  $5\%$  for all values of $\beta$ considered. This is particularly impressive considering that at the lowest temperature there are 3 orders of magnitude difference between the classical and quantum rates. The LGR rate is in fact slightly more accurate at this temperature than the GR-QTST rate, which has an error of about $8\%$ at $\beta=6$. \cite{Thapa19} 
 
\subsection{Multi-dimensional Spin Boson Model}
The second system we shall consider is a multi-dimensional spin-boson model,
\begin{subequations}
\begin{align}
V_0(\bm{q}) &= \sum_{\nu=1}^f \frac{1}{2}m\omega_\nu^2 q_\nu^2 + c_\nu q_\nu \\
V_1(\bm{q}) &= \sum_{\nu=1}^f \frac{1}{2}m\omega_\nu^2 q_\nu^2 - c_\nu q_\nu -\epsilon,
\end{align}
\end{subequations}
in which the bosonic bath modes are described by an exponentially damped Ohmic spectral density 
\begin{equation}
J(\omega) = \Lambda \frac{\pi\omega}{4\omega_c} e^{-\omega/\omega_c}, \label{exp_damped_ohmic}
\end{equation}
 with a reorganisation energy of $\Lambda=50\,\mathrm{kcal}\,\mathrm{mol}^{-1}$ and a cutoff frequency of $\beta\hbar\omega_c=8$ at $300\,\mathrm{K}$.  
 
This model was chosen to be strongly quantum mechanical, so as to provide a stringent test of the accuracy of GR-QTST and LGR. The exact quantum mechanical golden-rule rate can be calculated for comparison by numerical integration of the exact expression for $c(t)$\cite{Weiss08}
\begin{equation}
\frac{c(t)}{Q_r} = \exp(-i\epsilon t/\hbar - \phi(t)/\hbar)  \label{Exact_SB_rate}
\end{equation}
where
\begin{equation}
\phi(t) = \frac{4}{\pi}\int\frac{J(\omega)}{\omega^2}\Bigg[\frac{1-\cos(\omega t)}{\tanh(\beta\hbar\omega/2)}-i \sin(\omega t)\Bigg]\mathrm{d}\omega,
\end{equation}
and the classical limit of the rate is given by Marcus theory \cite{Marcus56,Marcus85,Hush61,Hush99}
\begin{equation}
k_{\mathrm{MT}}= \frac{\Delta^2}{\hbar}\sqrt{\frac{\pi\beta}{\Lambda}}e^{-\beta(\Lambda-\epsilon)^2/4\Lambda}. \label{Marcus_Theory}
\end{equation}

In order to calculate the exact, GR-QTST, and LGR rates for this problem we use a discretised form of the spectral density with $f$ modes, which in the limit as $f\to\infty$ becomes equivalent to the continuous form:
\begin{equation}
J(\omega)=\frac{\pi}{2}\sum_{\nu=1}^{f}\frac{c_{\nu}^2}{m\omega_\nu}\delta(\omega-\omega_\nu).
\end{equation}
The discretisation we employ is defined by
\begin{equation}
\omega_\nu = -\omega_c \ln(x_\nu),
\end{equation}
and
\begin{equation}
c_\nu = \omega_\nu \sqrt{\frac{m\Lambda w_\nu}{2} },
\end{equation}
where $w_\nu$ and $x_\nu$ are the weights and nodes of an $f$-point Gauss-Legendre quadrature on the interval $[0,1]$. Discretising using Gauss-Legendre quadrature rather than the more commonly used midpoint rule leads to more rapid convergence of the exact golden-rule rate with respect to the number of bath modes.  Since GR-QTST is not size consistent the GR-QTST rates do not converge with increasing $f$, and by using a rapidly convergent discretisation we will be able to more clearly illustrate the breakdown of GR-QTST with increasing system size. 

 \begin{figure}[t]
 \resizebox{1.0\columnwidth}{!} {\includegraphics{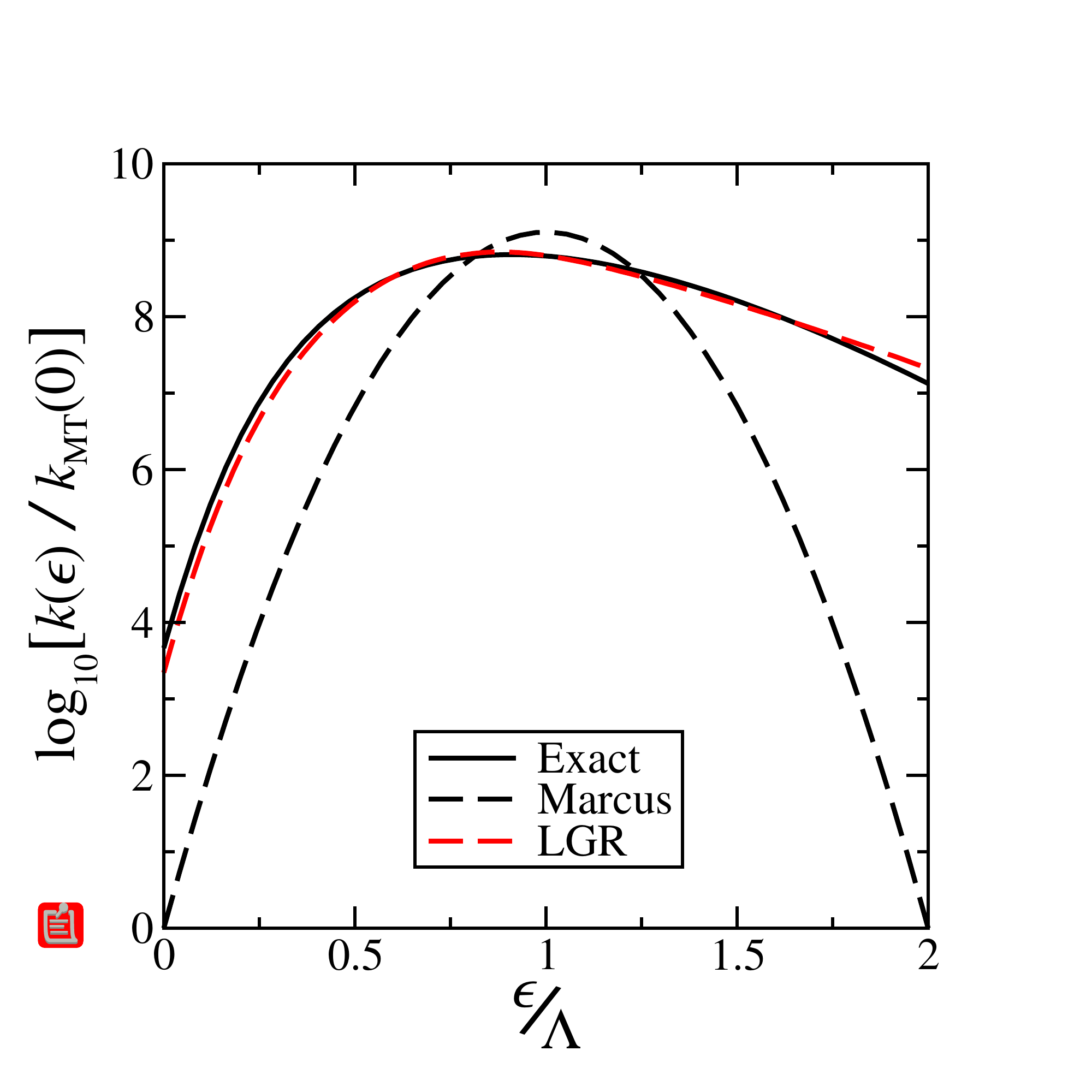}}
 \centering
 \caption{Exact, LGR and Marcus theory reaction rate constants as a function of the driving force for a multi-dimensional spin-boson model. All three rate constants are plotted relative to the classical (Marcus theory) rate at zero driving force ($\epsilon=0$). The LGR rates were computed using 256 path-integral beads and the exact rate was calculated by numerical integration of Eq.~(\ref{Exact_SB_rate}). Both the exact and LGR rates are converged to graphical accuracy with $f=16$ bath modes.}
 \label{SB_Rates}
 \end{figure}

\begin{figure}[t]
 \resizebox{1.0\columnwidth}{!} {\includegraphics{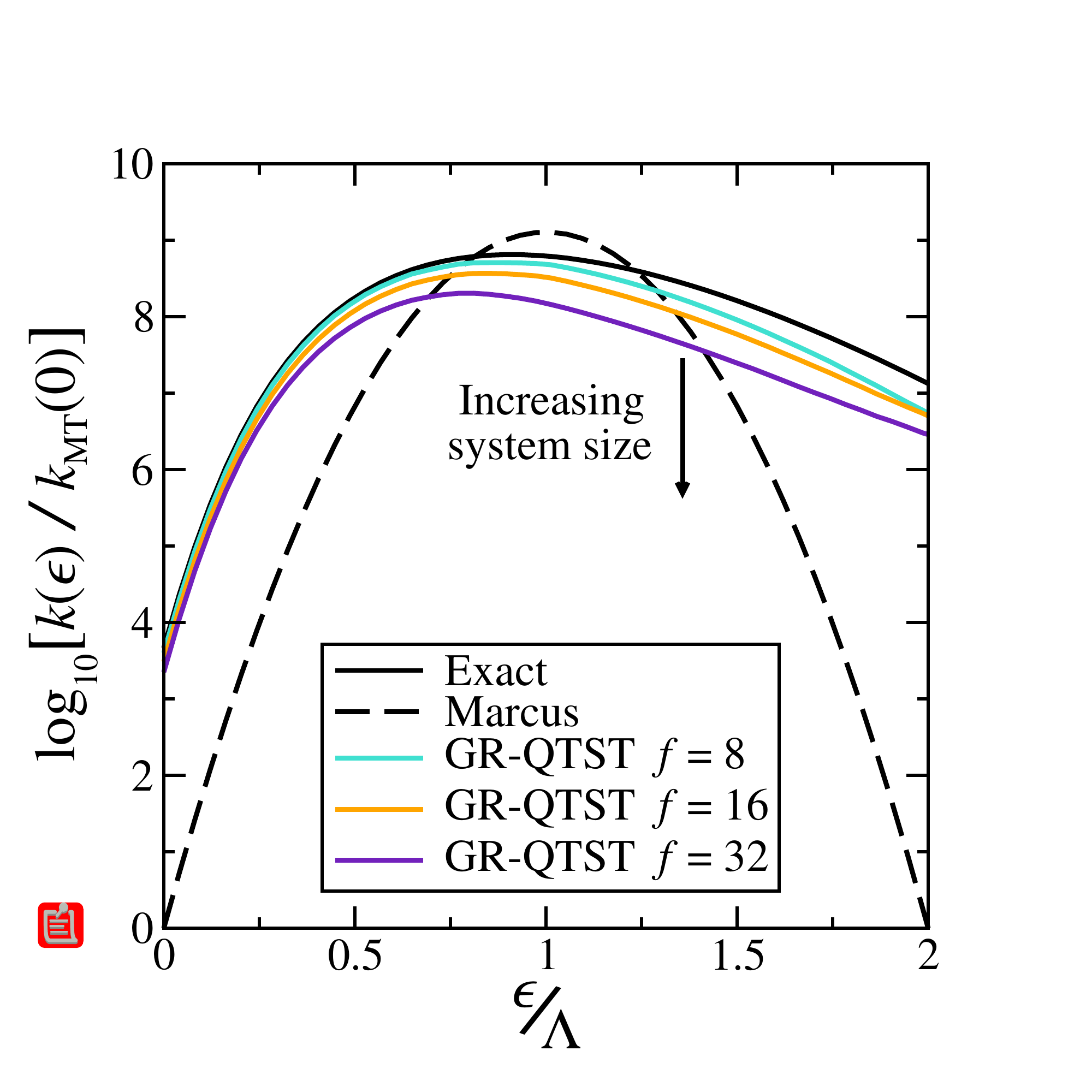}}
 \centering
 \caption{Comparison of GR-QTST reaction rate constants as a function of the driving force for the spin-boson model with an increasing number of bath modes, relative to the classical (Marcus theory) rate at zero driving force ($\epsilon=0$). The GR-QTST rates were computed using 256 path-integral beads and the exact rate was calculated by numerical integration of Eq.~(\ref{Exact_SB_rate}). This plot illustrates that GR-QTST does not converge as the size of the system increases, due to the lack of size consistency discussed in Sec.~III.B. }
 \label{SB_Rates_GRQTST_Breakdown}
 \end{figure}

Figure \ref{SB_Rates} shows the exact, LGR and classical (Marcus theory) rates as a function of driving force for this model, relative to the classical rate at zero driving force. We find that both the exact rates and the LGR rates are essentially converged to graphical accuracy using $f=8$ bath modes, and that they are very well converged with the $f=16$ modes that we use in this figure. Comparison of the classical rate and the exact quantum mechanical rate highlights the importance of nuclear quantum effects. We see that the exact rate is well over 3 orders of magnitude larger than the classical rate at $\epsilon=0$, and 7 orders of magnitude larger at $\epsilon=2\Lambda$. While the logarithm of the classical rate exhibits a famous parabolic dependence on the driving force, the exact rate is not symmetric about $\epsilon=\Lambda$. This asymmetry of the quantum rate as a function of driving force is well known, and occurs due to the increased efficiency of tunnelling in the Marcus inverted regime. We also note that the large decrease in the quantum compared to the classical rate near the activationless reaction, $\epsilon=\Lambda$, is an indication of the importance of high frequency modes in this system. It can be attributed to zero-point energy broadening the quantum distribution relative to the classical distribution at the same temperature, resulting in a reduced probability of the system being found at the diabatic crossing point.

We see that LGR reproduces the main qualitative features of the exact rate as a function of driving force.  The largest errors exhibited by LGR are at $\epsilon=0$, where it is just over a factor of 2 smaller than the exact rate, and at $\epsilon=2\Lambda$ where it is just under a factor of 2 larger than the exact rate. Considering the large difference between the classical and quantum rates for these systems, we feel that these are not unreasonable errors. Encouragingly LGR is also most accurate for values of $\epsilon$ where the difference between the quantum and classical rates is smallest, showing an error of less than $10\%$ at $\epsilon=\Lambda$. The accuracy of LGR for this system even in the inverted regime, where $\lambda_{\mathrm{sp}}<0$ and the rate must be evaluated at $\lambda=0$, is perhaps the most encouraging feature of Fig.~\ref{SB_Rates}. Especially since this calculation avoids the numerical analytic continuation needed to apply Wolynes theory to the inverted regime.\cite{Lawrence18} We would conclude from Fig.~\ref{SB_Rates} that the LGR provides an accurate approximation to quantum mechanical golden-rule rates for condensed phase systems in both the normal and inverted regimes. 

Figure \ref{SB_Rates_GRQTST_Breakdown} shows the GR-QTST results for discretisations with $f=8,16\text{ and }32$ bath modes, illustrating the failure of this method to converge with increasing system size. While GR-QTST provides reasonably accurate results for the 8 mode discretisation, as we move to 16 and 32 mode discretisations the error in the GR-QTST rate grows significantly. This is in stark contrast to both the exact and LGR rates which show no significant change with increasing system size. We see that the size consistency error is most pronounced in the inverted regime, where the GR-QTST rate is out by as much as a factor of 3 for $f=16$ and by almost an order of magnitude for $f=32$. Although the error in the normal regime is not as pronounced, it is important to stress that since the GR-QTST rates do not converge as $f\to\infty$ it is possible to obtain an arbitrarily large error by going to sufficiently large $f$. We shall demonstrate in a forthcoming paper that, for more realistic atomistic models of condensed phase electron transfer with thousands of degrees of freedom rather than only 32, the error even in the normal regime can become very significant.\cite{Lawrence20c}
 
 \section{Conclusion and future work} \label{LGR_Conclusion}
 
In this paper, we have proposed an alternative to Wolynes theory for calculating non-adiabatic reaction rates in the golden-rule limit. This alternative, which we have called the \lq\lq Linear Golden-Rule'' approximation, improves on the recent \lq\lq Golden-Rule Quantum Transition State Theory" approximation of Richardson and coworkers by eliminating its size consistency issues. Like GR-QTST, but unlike Wolynes theory, the method we have proposed recovers the correct classical golden-rule result in the high-temperature limit, and it is exact at all temperatures for the special case of two linearly crossing potentials in one dimension. Its advantage over GR-QTST is that it is size consistent, and can therefore be used without issue in large-scale simulations of condensed phase reactions: the uncoupled \lq\lq spectator" modes that are not involved in the reaction do not contribute to the calculated reaction rate.  Its advantages over Wolynes theory are threefold: it is exact in the high-temperature limit, it can be applied to reactions with more than one transition state, and it can be used to calculate electron transfer rates in the Marcus inverted regime without the need for any numerical analytic continuation.

In the previous section we have demonstrated the accuracy of LGR for both a one dimensional anharmonic pre-dissociation model and a multidimensional spin-boson model.
The approximate independence of the LGR approximation to the rate on $\lambda$ allows for direct evaluation of reaction rates in the Marcus inverted regime. It also means that LGR can be applied to systems with multiple transition states, which when treated separately would have different values of $\lambda_{\mathrm{sp}}$. Application of LGR to System I from Ref.~\citenum{Fang19} at $\beta=3$ reduced units, which was designed to exhibit two transition states, confirms the accuracy of LGR for such systems, with $k_{\mathrm{LGR}}/\Delta^2=2.2\pm0.2\times10^{-28}$ reduced units compared to $k/\Delta^2=1.98\times10^{-28}$ for the exact golden-rule rate, $k_{\mathrm{cl}}/\Delta^2=1.1\pm0.2\times10^{-29}$ for the classical golden-rule rate and $k_{\text{GR-QTST}}/\Delta^2=2.3\pm0.2\times10^{-28}$ for the GR-QTST rate.\cite{Fang19} In contrast, Wolynes theory overestimates the rate by more than a factor of 300 for this system.\cite{Fang19} Although LGR eliminates the size inconsistency seen in GR-QTST, it also loses the connection to the semiclassical instanton.\cite{Richardson15a,Richardson15c,Heller20,Thapa19} As such, in low dimensional models, LGR is expected to become less accurate than GR-QTST at sufficiently low temperatures. However, we feel that for simulations of condensed phase reactions, such as electron transfer in solution, size consistency is more important than a formal connection to the semiclassical instanton. It may of course be possible to develop a method which is both size consistent and has a close connection to the semiclassical instanton, and this is an interesting avenue for future work.

We have recently shown that Wolynes theory can be generalised to calculate reaction rates beyond the golden-rule limit, to give a non-adiabatic quantum instanton approximation,\cite{Lawrence20b} which reduces to the projected quantum instanton in the adiabatic limit.\cite{Miller03,Vanicek05,Vaillant19}  The development of a generalisation of LGR, capable of treating systems with arbitrary couplings, is an important target of future work. In particular, one might speculatively hope that this would provide further insight into the development of an accurate non-adiabatic generalisation of RPMD, which is an area of active research in the field.\cite{Shushkov12,Richardson13,Ananth13,Duke16,Menzeleev14,Kretchmer16,Chowdhury17,Kretchmer18,Tao18,Tao19,Ghosh20,Lawrence19b} However, we note that, LGR can already be used to calculate reaction rates in systems with arbitrary coupling strengths, by combining it with Born-Oppenheimer RPMD using a simple interpolation formula which interpolates between the golden-rule and Born-Oppenheimer limits.\cite{Lawrence19a}  Future work will look to investigate the accuracy of this approach in systems where Wolynes theory is known to break down in the golden-rule limit. 

For now, LGR provides an accurate approach to calculating reaction rates in the golden-rule limit, which is straightforward to apply to condensed phase systems in both the normal and inverted regimes. In a forthcoming paper,\cite{Lawrence20c} we shall apply it to an atomistic model of aqueous ferrous-ferric electron transfer, and compare the results to those of Wolynes theory and GR-QTST. This will allow us to assess the recent suggestion that ferrous-ferric electron transfer exhibits a range of qualitatively different tunnelling pathways, leading to a break down of Wolynes theory and also the assumptions of linear response inherent in the Marcus picture of electron transfer.\cite{Fang20} The conclusions of this study will turn out to be entirely consistent with what we have found here.

 \acknowledgments
We are grateful to Lachlan Lindoy and Thomas Fay for helpful discussions, and to Jeremy Richardson for his comments on this manuscript. J. E. Lawrence is supported by The Queen's College Cyril and Phillis Long Scholarship in conjunction with the Clarendon Fund of the University of Oxford and by the EPRSC Centre for Doctoral Training in Theory and Modelling in the Chemical Sciences, EPSRC grant no. EP/L015722/1.

\section*{Data availability statement}
The data that support the findings of this study are available in the paper itself.

\appendix
\section{Ring Polymer Discretisation}
For completeness we give here the $n$ bead discretisation of Eqs.~(\ref{diabatic_difference_functional})-(\ref{Correction_terms}).  The diabatic energy gap averaged over the two bridging beads is simply
\begin{equation}
\bar{V}_{-,n}^{(l)}(\mathbf{q}) = \frac{V_-(\bm{q}_0)+V_-(\bm{q}_l)}{2}, 
\end{equation}
\color{black}
and the correction term is given by 
\begin{equation}
\bar{\mathcal{K}}_{-,n}^{(l)}(\mathbf{q})=\bar{\mathcal{K}}_{0,n}^{(l)}(\mathbf{q})-\bar{\mathcal{K}}_{1,n}^{(l)}(\mathbf{q})
\end{equation}
where 
\begin{equation}
\bar{\mathcal{K}}_{i,n}^{(l)}(\mathbf{q}) = \frac{\mathcal{K}_{i,n}^{(l)}(\mathbf{q},\bm{q}_0)+\mathcal{K}_{i,n}^{(l)}(\mathbf{q},\bm{q}_l) }{2}
\end{equation}
where, for $l=1,\dots,n-1$, 
\begin{subequations} 
\begin{align}
 \mathcal{K}^{(l)}_{0,n}(\mathbf{q},\bm{q}_j)&= \sum_{k=l}^n w_{kl} \frac{\bar{\bm{\kappa}}_0(\bm{q}_k,\bm{q}_j)\dotproduct\big(\bm{q}_k-\bm{q}_j\big)}{(n-l)}  \\
 \mathcal{K}^{(l)}_{1,n}(\mathbf{q},\bm{q}_j)&= \sum_{k=0}^l w_{kl}\frac{\bar{\bm{\kappa}}_1(\bm{q}_k,\bm{q}_j)\dotproduct\big(\bm{q}_k-\bm{q}_j\big)}{l}, 
\end{align} 
\end{subequations}
and $w_{kl}$ are the weights in Eq.~(13). Here the effective diabatic gradients are
\begin{equation}
\bar{\bm{\kappa}}_i(\bm{q}_k,\bm{q}_j) =  \frac{\big(\nabla V_i(\bm{q}_k) +\nabla V_i(\bm{q}_j)\big)\dotproduct\nabla V_{-}(\bm{q}_j)}{2|\nabla V_{-}(\bm{q}_j)|^2} \nabla V_{-}(\bm{q}_j).
\end{equation}
The end points $l=0$ ($\lambda_0=0$) and $l=n$ ($\lambda_n=\beta$) are special cases, for which   
\begin{subequations} 
\begin{align}
 \bar{\mathcal{K}}^{(l)}_{0,n}(\mathbf{q})&= \sum_{k=0}^{n-1} \frac{\bar{\bm{\kappa}}_0(\bm{q}_k,\bm{q}_0)\dotproduct\big(\bm{q}_k-\bm{q}_0\big)}{n}  \\
 \bar{\mathcal{K}}^{(l)}_{1,n}(\mathbf{q})&=0, 
\end{align} 
\end{subequations}
and 
\begin{subequations} 
\begin{align}
 \bar{\mathcal{K}}^{(l)}_{0,n}(\mathbf{q})&=0 \\
 \mathcal{K}^{(l)}_{1,n}(\mathbf{q})&=\sum_{k=0}^{n-1} \frac{\bar{\bm{\kappa}}_1(\bm{q}_k,\bm{q}_0)\dotproduct\big(\bm{q}_k-\bm{q}_0\big)}{n}, 
\end{align} 
\end{subequations}
respectively.

\end{document}